\documentclass[a4paper]{PoS}

\title{Forbidden Kaon and Pion Decays in NA62}

\ShortTitle{Forbidden Kaon and Pion Decays in NA62}

\author{\speaker{Matthew Moulson} for the NA62 Collaboration%
        \thanks{
G.~Aglieri Rinella, F.~Ambrosino, B.~Angelucci, A.~Antonelli, G.~Anzivino, 
R.~Arcidiacono, I.~Azhinenko, S.~Balev, A.~Biagioni, C.~Biino
% ...
}\\
       INFN Laboratori Nazionali di Frascati, Italy\\
       E-mail: \email{moulson@lnf.infn.it}}

\abstract{
NA62, an experiment at the CERN SPS to measure the branching ratio for the 
decay $K^+\to\pi^+\nu\bar{\nu}$ with $\sim$10\% precision, will observe
$\sim10^{13}$ $K^+$ decays in its fiducial volume, and will thus
also be able to carry out a rich program to search for $K^+$ and $\pi^0$ 
decays that are forbidden in the Standard Model, including in particular 
$K^+$ decays that violate the conservation of lepton flavor and/or number. 
NA62's potential performance in searches for a number of forbidden 
$K^+$ and $\pi^0$ decays is discussed, with initial sensitivity 
estimates.}

\FullConference{2013 Kaon Physics International Conference\\
		 Apr 29-30 and May 1st\\
		 University of Michigan, Ann Arbor, Michigan - USA}

\begin{document}

\section{Introduction}
The goal of the NA62 experiment at the CERN SPS \cite{NA62+10:TDD} 
is to measure the branching ratio (BR) for the decay 
$K^+\to\pi^+\nu\bar{\nu}$ with a precision of $\sim$10\%. 
The Standard Model (SM) prediction 
${\rm BR}(K^\pm\to\pi^\pm\nu\bar{\nu}) = (7.8\pm0.8)\times10^{-11}$ 
\cite{BGS11:Kpnn} is quite precise, due in large part to the fact that 
the hadronic matrix element for the decay can be obtained from
the experimentally determined rate for $K_{e3}$ decays. The value of 
${\rm BR}(K^\pm\to\pi^\pm\nu\bar{\nu})$ is 
therefore a sensitive probe for physics beyond the SM, but since this BR
is so tiny, prodigious numbers of kaons are required for its measurement. 
NA62 is expected to begin running in late 2014; about $10^{13}$ 
$K^+$s will decay inside the NA62 fiducial volume in two years' equivalent
of data taking. Obviously, this large sample of kaon decays provides an 
opportunity to perform various searches for novel phenomena in addition to 
$K^+\to\pi^+\nu\bar{\nu}$. In particular, NA62 should be well positioned to 
obtain competitive results on kaon decays with explicit violation of the 
conservation of lepton flavor or number (LFNV), as well as in related areas,
such as the search for heavy neutrinos in decays such as $K_{\mu2}$. In 
addition, since the decay mode $K^+\to\pi^+\pi^0$ accounts for $\sim$21\% 
of $K^+$ decays, and since the detection of the $\pi^+$
monochromatic in the $K^+$ rest frame effectively tags the $\pi^0$ in 
such decays, NA62 should also be able to improve on searches for new physics 
in a number of $\pi^0$ decays as well.

\section{Violation of lepton flavor and/or number conservation in kaon decays}
Many attempts to improve upon the SM introduce new interactions that may
give rise to violation of lepton flavor and/or number conservation in 
specific processes, including supersymmetry \cite{BHS95:SUSYLFV,H+96:SUSYLFV}, 
mechanisms for dynamical electroweak symmetry breaking with strong coupling
such as extended technicolor \cite{A+04:ETC}, 
Little Higgs models \cite{C+07:LHTLFV}, 
models that introduce heavy neutrinos into the SM \cite{BRS09:nuMSM}, 
models featuring large extra dimensions 
\cite{CN05:extraDimLFV,ABP06:extraDimLFV}, and more.
In the past, searches for LFNV in kaon decays have been able to place 
tight constraints on the parameter space for some of these models, 
for some very straightforward reasons. 
The availability of intense kaon beams has made it 
possible to design high-statistics experiments, while the relative
topological simplicity of kaon decays (relatively few decay channels, 
low final-state multiplicities) and clear experimental signatures for 
the LFNV decays make efficient background rejection possible. 
As a result, kaon decay experiments have
reached sensitivities to 
branching ratios as low as $10^{-12}$, which, by simple dimensional arguments,
can provide access to mass scales upwards of 100~TeV in the search for 
new physics at tree level (e.g., a new gauge boson mediating the 
tree-level $s\to d\mu e$ transition) \cite{CH80:LFVtree}. 
Precisely because the results from searches for LFNV kaon decays up 
through the 1990s posed such stringent constraints on models such as 
technicolor, for the past decade or so, it appears to have been tacitly 
assumed that it would be very difficult to make any further progress with
kaon decays \cite{Lit05:rareK}. 
However, interest in searches for LFNV in 
charged lepton decays has remained robust, as witness the interest in 
experiments such as MEG and Mu2e as well as in searches 
for $\tau\to\mu\gamma$, for example, at next-generation flavor factories 
\cite{MMR08:CLFV}. In part this is because supersymmetric SM 
extensions that would explain the $>3\sigma$ discrepancy between 
the measured and predicted values of the muon anomaly, 
$a_\mu \equiv (g_\mu-2)/2$, predict some degree of charged LFV 
\cite{CM01:SUSYamu,CK01:SUSYLFV}. In any case, there is certainly 
no lack of theories predicting observable LFNV phenomena.
In this context, the next-generation experiments
planned to measure the BRs for decays such as $K^+\to\pi^+\nu\bar{\nu}$
represent real opportunities to push down the limits on LFNV phenomena through 
the study of charged kaon decays. Table~\ref{tab:modes} lists some 
LFNV $K^+$ decays 
(mainly those with three charged particles in the final state) and current 
limits on the corresponding BRs, as well as two of the best results 
obtained with $K_L$ decays, for comparison. The NA62 experiment at the CERN SPS 
should be able to significantly improve on many, if not all, of the limits
listed for the $K^+$ decays.
\begin{table}
\begin{center}
\begin{tabular}{lccc}
\hline\hline
Mode & UL at 90\% CL & Experiment & Ref. \\
\hline
$K^+ \to \pi^+\mu^+e^-$     & $1.3\times10^{-11}$  & BNL 777/865 & \cite{E865+05:Kpme} \\
$K^+ \to \pi^+\mu^-e^+$     & $5.2\times10^{-10}$  & BNL 865 & \cite{E865+00:KLFV} \\
$K^+ \to \pi^-\mu^+e^+$     & $5.0\times10^{-10}$  & BNL 865 & \cite{E865+00:KLFV} \\
$K^+ \to \pi^-e^+e^+$       & $6.4\times10^{-10}$  & BNL 865 & \cite{E865+00:KLFV} \\
$K^+ \to \pi^-\mu^+\mu^+$   & $1.1\times10^{-9}$   & NA48/2 & \cite{NA48+11:Kpmm} \\
$K^+ \to \mu^-\nu e^+e^+$   & $2.0 \times10^{-8}$  & Geneva-Saclay & \cite{S118+76:rare} \\
$K^+ \to e^-\nu \mu^+\mu^+$ & no data             & & \\
$K_L \to \mu e$             & $4.7\times10^{-12}$  & BNL 871 & \cite{E871+98:KLme} \\
$K_L \to \pi^0\mu e$        & $7.6\times10^{-11}$  & KTeV & \cite{KTeV+08:KLFV} \\
\hline\hline
\end{tabular}
\end{center}
\caption{Current status of searches for selected LFV and LNV $K^+$ decays 
for which limits can potentially be improved by NA62. Two of the best results 
obtained with $K_L$ decays are also listed for comparison.} 
\label{tab:modes}
\end{table}

\section{$K^\pm\to\pi^\mp\mu^\pm\mu^\pm$}
\begin{figure}
\begin{center}
\includegraphics[width=0.5\textwidth]{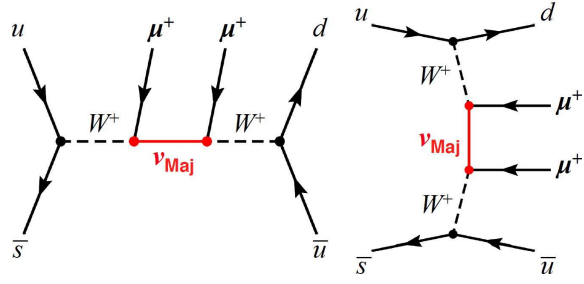}
\end{center}
\caption{Lowest-order diagrams contributing to $K^+\to\pi^-\mu^+\mu^+$.
The decay can proceed if the neutrino exchanged can annihilate itself, i.e.,
if it is its own antiparticle.}
\label{fig:Maj}
\end{figure}

The presence of the decay $K^+\to\pi^-\mu^+\mu^+$ in 
Tab.~\ref{tab:modes} is particularly interesting for two reasons. 
On the theoretical side, as discussed in \cite{LS00:KLNV,A+09:Maj}, the 
lepton number violation in this decay would imply that the virtual neutrino 
exchanged is a Majorana fermion, i.e., it is its own antiparticle, as 
illustrated in Fig.~\ref{fig:Maj}. This is similar to the case of 
neutrinoless nuclear double beta decay, and is an intriguing possibility, 
because if the neutrino is a Majorana 
fermion, the see-saw mechanism provides a natural explanation for the  
lightness of the observed $\nu_e$, $\nu_\mu$, and $\nu_\tau$ flavor 
eigenstates. In this scenario, the existence of heavy-neutrino mass 
eigenstates that participate in the neutrino mixing to form sterile, 
right-handed neutrino flavor eigenstates would also be predicted. 
On the experimental side, the decay 
$K^+\to\pi^-\mu^+\mu^+$ is interesting because NA62's prospects 
for measuring it can be accurately evaluated.  

Until recently, the most stringent limit on the branching ratio for this decay 
was that from the E865 decay-in-flight experiment at Brookhaven 
\cite{E865+00:KLFV}. This result was obtained
from the analysis of a few hundred candidate events reconstructed in the 
magnetic spectrometer. Most of these are actually $K_{\pi3}$ 
decays; analysis of the invariant-mass distribution with assigned particle
identification $M(\pi^-\mu^+\mu^+)$ gives 5 candidates in the signal window
around $m_{K^\pm}$ with an expected background (from sidebands) of 5.3 events.
This is used to set the limit ${\rm BR}(K^+\to\pi^-\mu^+\mu^+) <
3.0\times10^{-9}$ (90\% CL).

\begin{figure}
\begin{center}
\includegraphics[angle=270,width=0.80\textwidth]{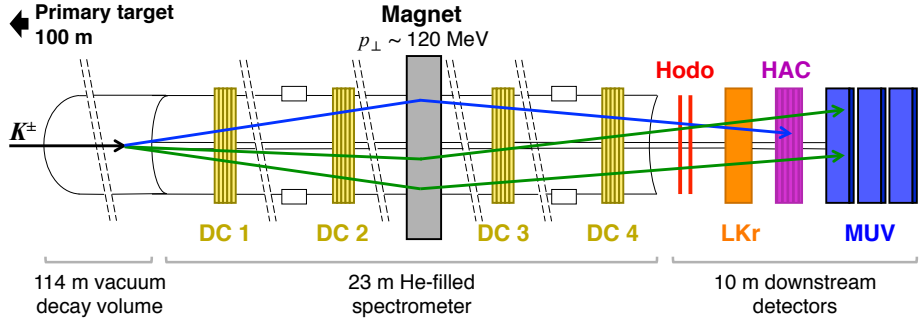}
\end{center}
\caption{Schematic diagram of the NA48/2 experiment, showing drift chambers (DC1--4), trigger hodoscope (Hodo), NA48 liquid-krypton electromagnetic calorimeter (LKr), hadronic calorimeter (HAC), and muon vetoes (MUV).}
\label{fig:NA48}
\end{figure}

More recently, NA48/2 performed a similar analysis. The primary purpose
of the NA48/2 experiment, which took data in 2003--2004, was to study direct 
$CP$ violation in the $K^+/K^-$ system. The experimental configuration is 
shown in Fig.~\ref{fig:NA48}. Simultaneous 60-GeV $K^+$ and $K^-$ beams 
entered a 114-m long vacuum decay tank, downstream of which, 
a 23-m long, helium-filled spectrometer consisting of four drift chambers
(DC1--4) and an analyzing magnet with a $p_\perp$ kick of 120~MeV was used 
to track and analyze charged secondaries. Downstream of the spectrometer 
were located a scintillator hodoscope (Hodo) to provide a fast trigger, 
the high-performance NA48 liquid-krypton electromagnetic calorimeter (LKr), 
an iron/scintillator hadronic calorimeter (HAC), and a stack of 
muon-veto detectors (MUV). The $K\to\pi\mu\mu$ (signal) and 
$K\to\pi\pi\pi$ (normalization) samples were both collected with the same
two-level trigger for three-track decays.
For both signal and normalization events, reconstruction of the three-track
vertex with strict quality criteria was required. For signal events, two 
tracks were further required to be identified as muons by hits in the MUV.
The experimental configuration and the analysis are further described 
in \cite{NA48+11:Kpmm}.

\begin{figure}
\begin{center}
\includegraphics[width=0.75\textwidth]{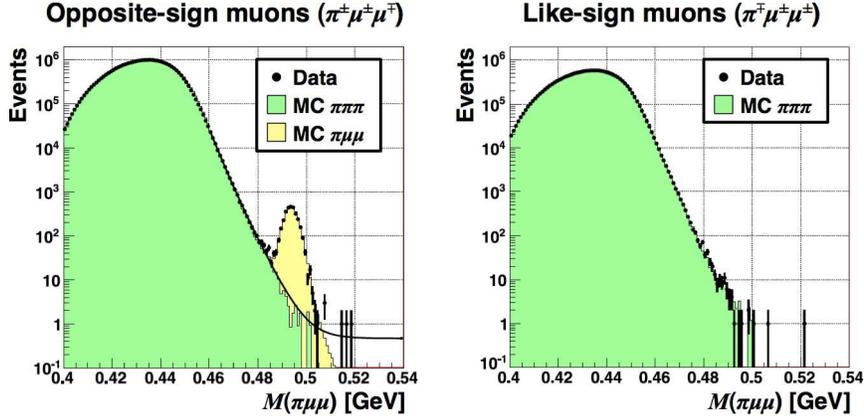}
\end{center}
\caption{NA48/2 invariant-mass distributions for $K\to\pi\mu\mu$ candidates.}
\label{fig:Kpmm}
\end{figure}

Figure~\ref{fig:Kpmm} shows the NA48/2 invariant-mass distributions for 
$K\to\pi\mu\mu$ candidate events with opposite-sign muons (left) and same-sign 
muons (right). The decay with opposite-sign muons 
($K^\pm\to\pi^\pm\mu^\pm\mu^\mp$) conserves lepton number; 
it is a flavor-changing neutral-current decay for which there are
form-factor predictions from chiral perturbation theory. NA48/2 sees about 
3000 signal candidates, visible as the peak near $M(\pi\mu\mu) = m_{K^\pm}$
in Fig.~\ref{fig:Kpmm}, left. Normalized Monte Carlo (MC) distributions 
for signal and background are shown as the yellow- and green-shaded regions.
In the distribution for events with like-sign muons, which violate lepton number
conservation, the signal peak is largely absent (Fig.~\ref{fig:Kpmm}, right).
There are 52 signal candidates in the region near $M(\pi\mu\mu) = m_{K^\pm}$ 
and $52.6\pm19.8$ background events expected from MC. This gives the upper 
limit ${\rm BR}(K^\pm\to\pi^\mp\mu^\pm\mu^\pm) < 1.1\times10^{-9}$ (90\% CL), 
which improves upon the E865 result by about a factor of three.

This analysis, however, was not fully optimized for the purposes of rejecting
$K_{\pi3}$ background to $K^\pm\to\pi^\mp\mu^\pm\mu^\pm$. Subsequent analysis 
demonstrated that the $K_{\pi3}$ events with $M(\pi\mu\mu) \approx m_{K^\pm}$ 
are all of the type where two of the pions decay to muons, and one of the
$\pi\to\mu$ decays occurs downstream of the spectrometer magnet and upstream
of the last drift chamber. In this topology, the latter track (which is 
identified as a muon) is misreconstructed; this can increase the 
apparent value of $M(\pi\mu\mu)$. Requiring that the pion track have 
$p > 20$~GeV decreases the acceptance for signal events by 50\%, but
also eliminates all $K_{\pi3}$ events with $M(\pi\mu\mu) \approx m_{K^\pm}$.
With the background thus reduced by at least an order of magnitude, 
the NA48/2 sensitivity is increased to $\sim$10$^{-10}$. 

\section{NA62 prospects for searches for LFNV kaon decays}
\label{sec:LFNV}

\begin{figure}
\begin{center}
\includegraphics[angle=270,width=0.80\textwidth]{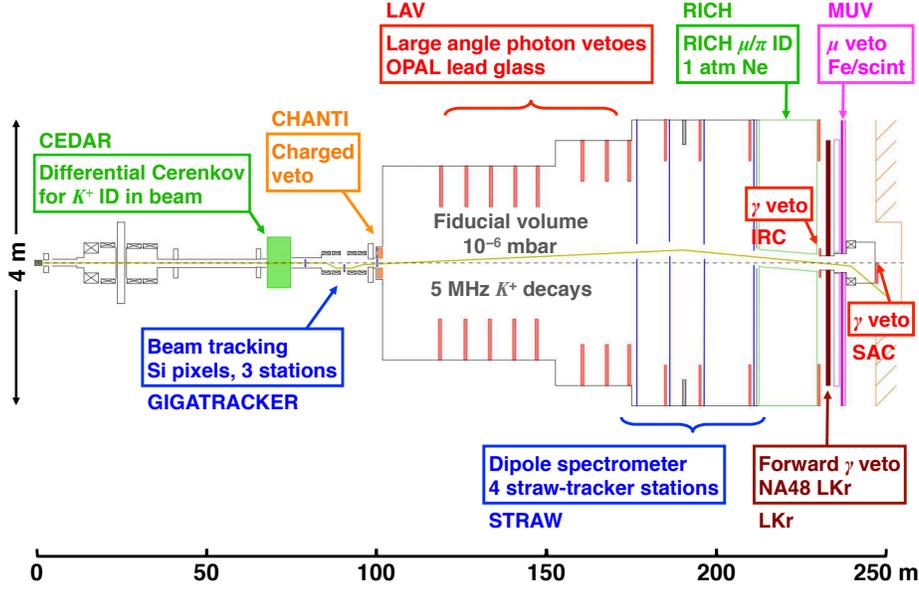}
\end{center}
\caption{Schematic diagram of the NA62 experiment.}
\label{fig:NA62}
\end{figure}

NA62 should be able to do better still. The experimental setup is illustrated 
schematically in Fig.~\ref{fig:NA62}. The basic infrastructure of the NA62 
experiment is inherited from NA48, including some of the beamline 
elements, most of the vacuum tank, and most importantly, the LKr calorimeter, 
which will be used as a high-performance veto for forward photons in the 
measurement of ${\rm BR}(K^+\to\pi^+\nu\bar{\nu})$. Most of the other 
detectors are either new or have been rebuilt.
The current status of NA62 installation is summarized in \cite{Hah13:Kaon}.

The experiment makes use of an unseparated, 
750-MHz, 75-GeV positive beam that is $\sim$6\% $K^+$. The decay region, 
of fiducial length $\sim$65 m for the study of $K^+\to\pi^+\nu\bar{\nu}$, 
is evacuated to $10^{-6}$~mbar. Inside the fiducial volume, there are 
$\sim$5~MHz of $K^+$ decays. With $1\times10^{6}$ live seconds per year 
(corresponding to 100 days of data taking, accounting for effective duty 
factor, uptime efficiency, etc.), in two years of data, there will be 
about $10^{13}$ $K^+$ decays in the NA62 fiducial volume.  
In addition to a 60-fold increase in statistics relative to NA48/2, 
NA62 will be able to take advantage of several of the 
upgrades critical to the success of the $K^+\to\pi^+\nu\bar{\nu}$ 
measurement in its program to search for LFNV kaon decays. Among these, 
the improvement in kinematic reconstruction for charged tracks from 
fast and precise tracking of individual beam particles (with the Gigatracker),
improved tracking for secondaries by four new straw chambers operated in 
vacuum, and a higher $p_\perp$ kick (270~MeV) from the spectrometer magnet
will lead to improved invariant-mass resolution for three-track vertices
(from 2--4~MeV to 1--2~MeV). Also important are NA62's redundant
particle-identification systems, including a new RICH and a completely
overhauled MUV system, which is highly segmented and included in 
the level-0 trigger.

In terms of the limit on ${\rm BR}(K^+\to\pi^-\mu^+\mu^+)$,
the better kinematic reconstruction in NA62 should enable significant 
improvements. In particular, simulations show that with the higher field 
and better three-track invariant-mass resolution expected in NA62, the 
tail towards higher values of $M(\pi\mu\mu)$ from $K_{\pi3}$ seen in
NA48/2 data (Fig.~\ref{fig:Kpmm}) is essentially eliminated, even without 
cuts on the momentum of the pion track. Thus it may be possible to reach 
the NA62 single-event sensitivity for $K^+\to\pi^-\mu^+\mu^+$, which 
is $10^{-12}$.

In a similar vein, preliminary studies have been performed for all of the 
decays listed in Tab.~\ref{tab:modes}, focused mainly on trigger strategies
(including rate and efficiency evaluation) and acceptance simulations.
The trigger criteria, particularly at level 0, are a natural concern, because
the use of a generic three-track trigger as in NA48/2 would consume a 
large part of the NA62 bandwidth (perhaps $\sim$600~kHz, compared to the 
$\sim$1~MHz design rate). Several primitives for the level-0 decision can 
be formed, including the following:
\begin{itemize}
\item ${\rm Q}_{\geq n}$: Hits in at least $n$ hodoscope quadrants
\item ${\rm LKR}_{\geq n}(x)$: At least $n$ clusters with energy greater than $x$ GeV in the
LKr
\item ${\rm MUV}_{\geq n}$: Hits in at least $n$ MUV pads
\end{itemize}
On the basis of these, triggers for lepton pairs can be implemented:
\begin{itemize}
\item $ee$ pairs: ${\rm Q}_{\geq 2} \cdot {\rm LKR}_{\geq 2}(15)$
\item $e\mu$ pairs: ${\rm Q}_{\geq 2} \cdot {\rm LKR}_{\geq 1}(15) \cdot {\rm MUV}_{\geq 1}$
\item $\mu\mu$ pairs: ${\rm Q}_{\geq 2} \cdot {\rm MUV}_{\geq 2}$
\end{itemize}
The exact criteria remain to be defined, of course, but studies indicate that 
all of the above triggers could be included at level 0 at the cost of a few
tens of kHz of rate. 
The NA62 acceptance (including the trigger efficiency) for each of the $K^+$ 
decays in Tab.~\ref{tab:modes} has been preliminarily evaluated by fast 
MC simulation.
As a general statement, the overall acceptances for final states containing
$ee$, $e\mu$, and $\mu\mu$ pairs are respectively a few percent, about 10\%, 
and about 20\%, with the LKr energy threshold for $e^\pm$ identification 
responsible for the lower efficiency for the channels containing $e^\pm$.
Considering the expected flux of $\sim$10$^{13}$ $K^+$s, the NA62 single-event 
sensitivities are $\sim$10$^{-12}$ for the $K^+$ decays.
Assuming that the backgrounds can be tamed, NA62 is well positioned to
improve significantly on the scenario in Tab.~\ref{tab:modes}.

\section{Searches for heavy neutrinos in $K_{\mu2}$ decay}

\begin{figure}
\begin{center}
\includegraphics[width=0.75\textwidth]{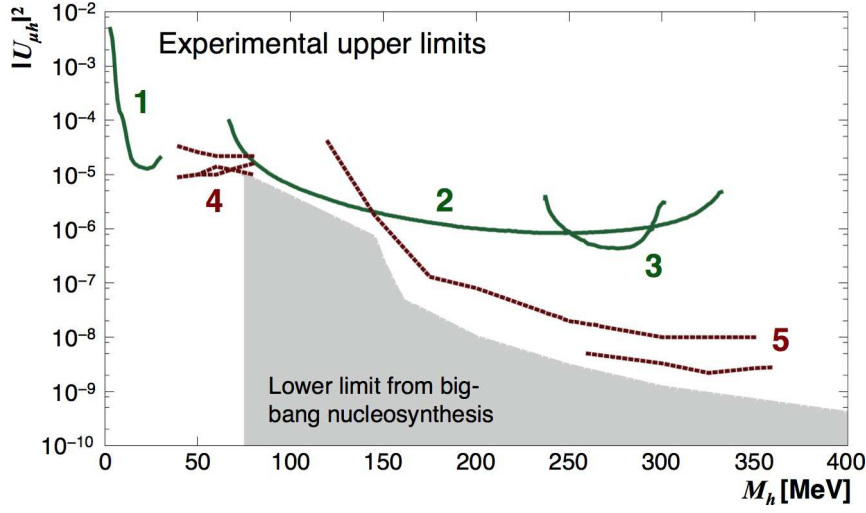}
\end{center}
\caption{Searches for heavy neutrinos in $K^\pm$ decays. Limits in green
are from production searches: 
1) PSI 1981, $\pi_{\mu2}$, \cite{A+81:hnu};
2) KEK 1982, $K_{\mu2}$, \cite{H+82:hnu};
3) LBL 1973, calculated from limits on $K\to\mu\nu+{\rm inv}$, \cite{P+73:hnu}.
Limits in red are from decay searches:
4) ISTRA+ 2012, $K_{\mu2}$ with $\nu_h\to\nu\gamma$, 
$\tau_h = 10^{-9}$, $10^{-10}$, and $10^{-11}$~s and Majorana statistics assumed 
\cite{ISTRA+12:hnu};
5) PS191 1988, $K_{\mu2}$ with $\nu_h\to\mu^\mp e^\pm\nu$ (above), 
$\nu_h\to\mu^\mp \pi^\pm$ (below) \cite{PS191+88:hnu}.
The gray shaded region indicates a {\em lower} limit obtained from models for
big-bang nucleosynthesis.}
\label{fig:hnu}
\end{figure}

A related topic is the direct search for heavy-neutrino mass eigenstates
$\nu_h$ among the decay products of dominant $K^\pm$ decays with final-state 
neutrinos.
There are two types of searches. In production searches, one analyzes
the momentum spectrum of a decay such as $K_{\mu2}$ for kinematic evidence 
of the presence of a heavy particle. In decay searches, one attempts to 
exclusively reconstruct decays of the heavy neutrino itself in one or more 
hypothetical channels. Since decay searches focus on a particular topology, 
they usually obtain greater sensitivity, but only for the decay modes assumed.
In addition, decay searches assume that the lifetime of the heavy neutrino
is short enough so that the acceptance of the experiment for the decay 
of interest is not significantly affected. Since production searches are 
inclusive, the particular assumption on the lifetime is weaker, but it is 
generally assumed that the heavy neutrino will not decay inside the 
acceptance leaving signals that complicate the recognition of the parent 
decay (e.g., $K_{\mu2}$). The current status of searches for heavy neutrinos
in $K^\pm$ and $\pi^\pm$ decays is summarized in Fig.~\ref{fig:hnu}, which 
presents limits on the squared PMNS matrix element 
$|U_{\mu h}|^2$ that couples the heavy-neutrino mass eigenstate $\nu_h$ 
to the flavor eigenstate $\nu_\mu$ as a function of $m_h$, the mass of $\nu_h$.

NA62 has investigated the possibility of performing a production search for 
heavy neutrinos in 
$K_{\mu2}$ decay with 40\% of the 2007 $K_{\mu2}$ sample used for the 
measurement of $R_K = \Gamma(K_{e2})/\Gamma(K_{\mu2})$, as in \cite{NA62+11:RK},
corresponding to 18M $K_{\mu2}$ events. In the absence of backgrounds, this 
would allow upper limits to be set on $|U_{\mu h}|^2$ of $10^{-7}$ for 
$100<m_h<400$~MeV. The main backgrounds are from events with muons from 
beam pions and the beam halo, $K_{\pi2}$ and $K_{\pi3}$ decays with 
missed photons and $\pi\to\mu$ decays or a $\pi$ wrongly identified 
as a $\mu$, $K_{\mu2\gamma}$ and $K_{\mu3}$ decays with missed photons, 
and $K_{\mu2(\gamma)}$ decays with 
misreconstructed tracks. In the preliminary analysis, a single cleanly 
reconstructed track, identified as a $\mu$ on the basis of 
$E({\rm LKr})/p({\rm track})$ and confirmed by the 
MUV\footnote{Where possible---not all NA48 detectors were available during 
2007 running, as described in \cite{NA62+11:RK}.} is required. A veto is placed
on the presence of any other photons in the LKr, and halo cuts (in the plane
of track momentum vs.\ vertex position) are used to eliminate halo muons.
Systematics are still under evaluation, but the preliminary results indicate
that with the 2007 data alone, NA62 should be able to improve on the limits
from production searches shown in Fig.~\ref{fig:hnu} in the high-mass region 
($300 < m_h < 350$~MeV).
With the data from $K^+\to\pi^+\nu\bar{\nu}$ running, precise tracking 
(including beam tracking), and muon identification from the RICH and MUV 
in all data, NA62 should be able to improve on these results by orders 
of magnitude. 

\section{Searches for new physics in $\pi^0$ decays}

The use of kaon decays as a source of $\pi^0$'s to search for rare $\pi^0$ 
decays is well established. For example, KTeV used $K_L\to3\pi^0$ decays 
to search for several rare $\pi^0$ decays, while E787/E949 at Brookhaven used 
$K^+\to\pi^+\pi^0$ decays at rest as a source of tagged $\pi^0$'s in searches
for $\pi^0\to\gamma X$ \cite{E787+92:gX} and $\pi^0\to\nu\bar{\nu}$ (see below).
With $10^{13}$ $K^+$ decays in its fiducial volume from two years
of data, NA62 will have $\sim$$2.5\times10^{12}$ tagged $\pi^0$ decays 
from $K^+\to\pi^+\pi^0$ with which to conduct various searches.

For example, NA62 should be able to improve on limits for the decays 
$\pi^0\to 3\gamma$ and $\pi^0\to 4\gamma$. The former violates $C$-parity 
conservation and cannot be accommodated in the SM at any observable level. 
The latter has an SM branching ratio on the order of $10^{-11}$ due mainly
to higher-order electromagnetic contributions (light-by-light scattering)
\cite{BKS95:pi04g}; deviations from this prediction might provide evidence 
for new light scalars weakly coupled to the electromagnetic current. Current
experimental limits for both BRs are at the level of $10^{-8}$. At NA62, the 
primary challenge will be rejection of background from 
$K^+\to\pi^+\pi^0(\gamma)$ and $K^+\to\pi^+\pi^0\pi^0$, starting at the 
lowest trigger level. Studies indicate that a dedicated level-0 trigger 
such as ${\rm Q}_{=1}\cdot{\rm LKR}_{=3}(1)\cdot\overline{\rm MUV}$ can keep 
rates down to acceptable levels while maintaining reasonable efficiency, 
particularly if the trigger logic can exclude the $\pi^+$ cluster from the 
LKr cluster count.
At level-1, the trigger would have to make kinematic cuts on the 
LKr clusters (e.g., in the space of $M(\gamma\gamma\gamma)$ vs.\ $p_{\pi}^*$).
In the analysis (or possibly even at trigger level) the full complement 
of large- and small-angle photon veto detectors can be used to reject events
with additional photons, and a kinematic fit to the complete event can provide
a background rejection factor of $10^{-4}$. Then, NA62 should be able to obtain
BR limits at the level of $10^{-10}$, about two orders of magnitude better 
than present limits.   

\begin{figure}
\begin{center}
\includegraphics[angle=270,width=0.5\textwidth]{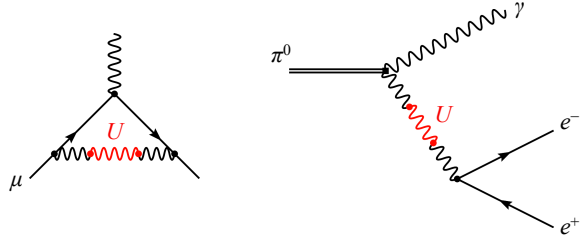}
\end{center}
\caption{Amplitudes involving the $U$ boson contributing to $a_{\mu}$ (left)
and to the decay $\pi^0\to e^+e^-\gamma$ (right).}
\label{fig:dark}
\end{figure}

NA62 is also well positioned to use $\pi^0$ decays to search for a new, 
light vector gauge boson with weak couplings to charged SM fermions, a 
so-called $U$ boson, or ``dark photon''. A hypothetical $U$ boson
could mediate the interactions of dark-matter constituents, as such 
providing explanations for various unexpected astrophysical observations 
and the results of certain dark-matter searches, and could also explain the 
$>3\sigma$ discrepancy between the measured and predicted values for 
the muon anomaly, $a_\mu$ (see \cite{PRV08:dark,Pos09:dark} and references
therein). A $U$ boson with a mass of less than $m_{\pi^0}/2$ might be directly 
observable in $\pi^0\to e^+e^-\gamma$ decays, as illustrated in 
Fig.~\ref{fig:dark}. The current upper limit on BR($\pi^0\to U\gamma$) with 
$U\to e^+e^-$ is from the WASA-at-COSY experiment, and decreases from 
$\sim$$1\times10^{-5}$ at $m_U = 30$~MeV to a little over $2\times10^{-6}$ 
at 100~MeV \cite{WASA+13:dark}. This result was obtained with 
$\sim$$5\times10^5$ $\pi^0\to e^+e^-\gamma$ events.
At NA62, with an $ee$ pair trigger like that described in Sec.~\ref{sec:LFNV},
$\sim$$10^8$ $\pi^0\to e^+e^-\gamma$ decays can be collected per year, with
a mass resolution for the $ee$ pair of $\sim$1~MeV (which can be improved 
by kinematic fitting), compared to several MeV in the case of WASA-at-COSY.
As a result, NA62 should be able to improve on the current limit by at least 
two orders of magntiude.

As a final example, NA62 should be able to improve on limits for the 
invisible decay of the $\pi^0$. The least exotic decay to an invisible
final state is $\pi^0\to\nu\bar{\nu}$. This is forbidden by angular-momentum
conservation if neutrinos are massless; for a massive neutrino $\nu$ of 
a given flavor and mass $m_\nu < m_{\pi^0}/2$ with standard coupling to the 
$Z$, the calculation of the decay rate is straightforward. 
The experimental signature $\pi^0\to{\rm invisible}$ could also arise 
from $\pi^0$ decays to other weakly interacting neutral states. 
The best direct experimental limit to date is
${\rm BR}(\pi^0\to{\rm invisible}) < 2.7 \times 10^{-7}$ (90\% CL), from
E949 at Brookhaven \cite{E949+05:p0inv}. Neutrino-mass limits translate into
limits on ${\rm BR}(\pi^0\to\nu\bar{\nu})$ (though not on BRs to other 
invisible final states). The best limit on $m_{\nu_\tau}$ (the neutrino mass
for which the limits are least stringent), $m_{\nu_\tau} < 18.2$~MeV 
\cite{ALEPH+98:nu_tau}, implies ${\rm BR}(\pi^0\to\nu\bar{\nu}) < 
5\times10^{-10}$ (95\% CL), while limits from astrophysics and cosmology
imply ${\rm BR}(\pi^0\to\nu\bar{\nu}) < 3\times10^{-13}$ 
\cite{Nat91:p0inv,LN91:p0inv}. 

Experimentally, the process $K^+\to\pi^+\pi^0$ with $\pi^0\to{\rm invisible}$
is very similar to $K^+\to\pi^+\nu\bar{\nu}$, with the important difference
that in the former case, the $\pi^+$ is monochromatic in the rest frame of 
the $K^+$. This means that there is no help from kinematics in identifying
$K^+\to\pi^+\pi^0$, $\pi^0\to\gamma\gamma$ with two lost photons---the limit on
${\rm BR}(\pi^0\to{\rm invisible})$ essentially depends on the performance 
of the photon vetoes. At level 0, NA62 could use the same trigger criteria 
as for $K^+\to\pi^+\nu\bar{\nu}$. Inclusion at level~0 of information from the 
photon vetoes (in particular the LAVs) is already under consideration;
for $\pi^0\to{\rm invisible}$ it might be important to keep the level-1 rate
under control as kinematic cuts to exclude $K^+\to\pi^+\pi^0$ events are 
loosened. The dominant contribution to the trigger rate is from
events with one photon overlapping with the $\pi^+$ on the LKr calorimeter
and one photon lost. With stringent track-quality cuts for the $\pi^+$ and
additional cuts in the $(p_{\pi^+}, \theta_{\pi^+})$ plane to deselect 
events with low-energy, large-angle photons, the $\pi^0$ rejection can be 
increased by perhaps a factor of ten with respect to the NA62 baseline
rejection of $10^{-8}$. Then, NA62 would have the potential to set a limit
on ${\rm BR}(\pi^0\to{\rm invisible})$ of $\sim$$10^{-9}$, which is about 100
times better than present limits.  

\section{Conclusions} 

Next-generation experiments to measure $K^+\to\pi^+\nu\bar{\nu}$ will be 
well adapted to carry out a rich program to search for very rare or forbidden
$K^+$ and $\pi^0$ decays, because intense $K^+$ sources and robust 
background rejection (through precise tracking and particle identification and 
hermetic photon vetoes) are the defining features of such experiments.
With $\sim$$10^{13}$ $K^+$ and $\sim$$2.5\times10^{12}$ $\pi^0$ decays in
its fiducial volume after two years' worth of data taking, NA62 will have 
single-event sensitivities of $\sim$$10^{-12}$ for a number of $K^+$ decays
that violate lepton flavor and/or number conservation, as well as the 
potential to improve existing limits in a variety of searches for related 
phenomena.   

\bibliographystyle{elsart-num}
\bibliography{kaon13_proc}

\begin{thebibliography}{10}
\expandafter\ifx\csname url\endcsname\relax
  \def\url#1{\texttt{#1}}\fi
\expandafter\ifx\csname urlprefix\endcsname\relax\def\urlprefix{URL }\fi

\bibitem{NA62+10:TDD}
F.~Hahn~(ed.), et~al., {NA62} technical design document, {NA62} Document 10-07,
  \url{http://cds.cern.ch/record/1404985} (2010).

\bibitem{BGS11:Kpnn}
J.~Brod, M.~Gorbahn, E.~Stamou, Phys.\ Rev.\ D 83 (2011) 034030.

\bibitem{BHS95:SUSYLFV}
R.~Barbieri, L.~Hall, A.~Strumia, Nucl.\ Phys.\ B 445 (1995) 219.

\bibitem{H+96:SUSYLFV}
J.~Hisano, et~al., Phys.\ Rev.\ D 53 (1996) 2442.

\bibitem{A+04:ETC}
T.~Appelquist, et~al., Phys.\ Rev.\ D 70 (2004) 093010.

\bibitem{C+07:LHTLFV}
S.~Choudhury, et~al., Phys.\ Rev.\ D 75 (2007) 055011.

\bibitem{BRS09:nuMSM}
A.~Boyarsky, O.~Ruchayskiy, M.~Shaposhnikov, Annu.\ Rev.\ Part.\ Nucl.\ Sci. 59
  (2009) 191.

\bibitem{CN05:extraDimLFV}
W.-F. Chang, J.~Ng, Phys.\ Rev.\ D 71 (2005) 053003.

\bibitem{ABP06:extraDimLFV}
K.~Agashe, A.~Blechman, F.~Petriello, Phys.\ Rev.\ D 74 (2006) 053011.

\bibitem{CH80:LFVtree}
R.~Cain, H.~Harari, Nucl.\ Phys.\ B 176 (1980) 135.

\bibitem{Lit05:rareK}
L.~Littenberg, hep-ex/0512044 (2005).

\bibitem{MMR08:CLFV}
W.~Marciano, T.~Mori, J.~Roney, Annu.\ Rev.\ Nucl.\ Part.\ Sci. 58 (2008) 315.

\bibitem{CM01:SUSYamu}
A.~Czarnecki, W.~Marciano, Phys.\ Rev.\ D 64 (2001) 013014.

\bibitem{CK01:SUSYLFV}
Z.~Chacko, G.~Kribs, Phys.\ Rev.\ D 64 (2001) 075015.

\bibitem{E865+05:Kpme}
A.~Sher, et~al., Phys.\ Rev.\ D 72 (2005) 012005.

\bibitem{E865+00:KLFV}
R.~Appel, et~al., Phys.\ Rev.\ Lett. 85 (2000) 2877.

\bibitem{NA48+11:Kpmm}
{NA48~Collaboration}, J.~Batley, et~al., Phys.\ Lett.\ B 697 (2011) 107.

\bibitem{S118+76:rare}
A.~Diamant-Berger, et~al., Phys.\ Lett.\ B 62 (1976) 485.

\bibitem{E871+98:KLme}
D.~Ambrose, et~al., Phys.\ Rev.\ Lett. 81 (1998) 5734.

\bibitem{KTeV+08:KLFV}
{KTeV~Collaboration}, E.~Abouzaid, et~al., Phys.\ Rev.\ Lett. 100 (2008)
  131803.

\bibitem{LS00:KLNV}
L.~Littenberg, R.~Shrock, Phys.\ Lett.\ B 491 (2000) 285.

\bibitem{A+09:Maj}
A.~Atre, et~al., JHEP 0905 (2009) 030.

\bibitem{Hah13:Kaon}
F.~{Hahn for the NA62 Collaboration}, in: Proc.\ 2013 Kaon Phys.\ Int.\ Conf.\
  ({KAON} '13), Ann Arbor MI, USA, 2013, {PoS(KAON13)031}.

\bibitem{A+81:hnu}
R.~Abele, Phys.\ Lett.\ B 105 (1981) 263.

\bibitem{H+82:hnu}
R.~Hayano, et~al., Phys.\ Rev.\ Lett. 49 (1982) 1305.

\bibitem{P+73:hnu}
C.~Pang, et~al., Phys.\ Rev.\ D 8 (1973) 1989.

\bibitem{ISTRA+12:hnu}
V.~Duk, et~al., Phys.\ Lett.\ B 710 (2012) 307.

\bibitem{PS191+88:hnu}
G.~Bernardi, et~al., Phys.\ Lett.\ B 203 (1988) 332.

\bibitem{NA62+11:RK}
{NA62~Collaboration}, C.~Lazzeroni, et~al., Phys.\ Lett.\ B 698 (2011) 105.

\bibitem{E787+92:gX}
M.~Atiya, et~al., Phys.\ Rev.\ Lett. 69 (1992) 733.

\bibitem{BKS95:pi04g}
E.~Bratkovskaya, E.~Kuraev, Z.~Silagadze, Phys.\ Lett.\ B 359 (1995) 217.

\bibitem{PRV08:dark}
M.~Pospelov, A.~Ritz, M.~Voloshin, Phys.\ Rev.\ D 78 (2008) 115012.

\bibitem{Pos09:dark}
M.~Pospelov, Phys.\ Rev.\ D 80 (2009) 095002.

\bibitem{WASA+13:dark}
{WASA-at-COSY Collaboration}, P.~Adlarson, et~al., arXiv:1304.0671 (2013).

\bibitem{E949+05:p0inv}
{E949~Collaboration}, A.~Artamonov, et~al., Phys.\ Rev.\ D 72 (2005) 091102(R).

\bibitem{ALEPH+98:nu_tau}
{ALEPH Collaboration}, R.~Barate, et~al., Eur.\ Phys.\ J.\ C 2 (1998) 395.

\bibitem{Nat91:p0inv}
A.~Natale, Phys.\ Lett.\ B 258 (1991) 227.

\bibitem{LN91:p0inv}
W.~Lam, K.-W. Ng, Phys.\ Rev.\ D 44 (1991) 3345.

\end{thebibliography}

\end{document}